\documentclass[12pt,a4paper]{article}
\pdfoutput=1
\usepackage{epsfig}
\usepackage{mathrsfs}
\usepackage{graphics}
\usepackage{amssymb}

\usepackage{tikz}
\usepackage{comment}
\usetikzlibrary{decorations.pathmorphing}

\usepackage{amsmath}
\usepackage{amsthm}
\usepackage{amsfonts}

\def\hybrid{\topmargin -20pt    \oddsidemargin 0pt
        \headheight 0pt \headsep 0pt
        \textwidth 6.25in       
        \textheight 9.5in       
        \marginparwidth .875in
        \parskip 5pt plus 1pt   \jot = 1.5ex}

\hybrid

\def\baselinestretch{1.2}

\catcode`\@=11

\def\marginnote#1{}
\newcount\hour
\newcount\minute
\newtoks\amorpm
\hour=\time\divide\hour by60
\minute=\time{\multiply\hour by60 \global\advance\minute by-\hour}
\edef\standardtime{{\ifnum\hour<12 \global\amorpm={am}%
        \else\global\amorpm={pm}\advance\hour by-12 \fi
        \ifnum\hour=0 \hour=12 \fi
        \number\hour:\ifnum\minute<10 0\fi\number\minute\the\amorpm}}
\edef\militarytime{\number\hour:\ifnum\minute<10 0\fi\number\minute}

\def\draftlabel#1{{\@bsphack\if@filesw {\let\thepage\relax
   \xdef\@gtempa{\write\@auxout{\string
      \newlabel{#1}{{\@currentlabel}{\thepage}}}}}\@gtempa
   \if@nobreak \ifvmode\nobreak\fi\fi\fi\@esphack}
        \gdef\@eqnlabel{#1}}
\def\@eqnlabel{}
\def\@vacuum{}
\def\draftmarginnote#1{\marginpar{\raggedright\scriptsize\tt#1}}

\def\draft{\oddsidemargin -.5truein
        \def\@oddfoot{\sl preliminary draft \hfil
        \rm\thepage\hfil\sl\today\quad\militarytime}
        \let\@evenfoot\@oddfoot \overfullrule 3pt
        \let\label=\draftlabel
        \let\marginnote=\draftmarginnote
   \def\@eqnnum{(\theequation)\rlap{\kern\marginparsep\tt\@eqnlabel}%
\global\let\@eqnlabel\@vacuum}  }

\def\preprint{\twocolumn\sloppy\flushbottom\parindent 2em
        \leftmargini 2em\leftmarginv .5em\leftmarginvi .5em
        \oddsidemargin -.5in    \evensidemargin -.5in
        \columnsep .4in \footheight 0pt
        \textwidth 10.in        \topmargin  -.4in
        \headheight 12pt \topskip .4in
        \textheight 6.9in \footskip 0pt
        \def\@oddhead{\thepage\hfil\addtocounter{page}{1}\thepage}
        \let\@evenhead\@oddhead \def\@oddfoot{} \def\@evenfoot{} }

\def\numberbysection{\@addtoreset{equation}{section}
        \def\theequation{\thesection.\arabic{equation}}}

\def\underline#1{\relax\ifmmode\@@underline#1\else
        $\@@underline{\hbox{#1}}$\relax\fi}

\def\titlepage{\@restonecolfalse\if@twocolumn\@restonecoltrue\onecolumn
     \else \newpage \fi \thispagestyle{empty}\c@page\z@
        \def\thefootnote{\fnsymbol{footnote}} }

\def\endtitlepage{\if@restonecol\twocolumn \else \newpage \fi
        \def\thefootnote{\arabic{footnote}}
        \setcounter{footnote}{0}}  

\catcode`@=12
\relax

\def\figcap{\section*{Figure Captions\markboth
        {FIGURECAPTIONS}{FIGURECAPTIONS}}\list
        {Figure \arabic{enumi}:\hfill}{\settowidth\labelwidth{Figure
999:}
        \leftmargin\labelwidth
        \advance\leftmargin\labelsep\usecounter{enumi}}}
 \relax
\def\tablecap{\section*{Table Captions\markboth
        {TABLECAPTIONS}{TABLECAPTIONS}}\list
        {Table \arabic{enumi}:\hfill}{\settowidth\labelwidth{Table
999:}
        \leftmargin\labelwidth
        \advance\leftmargin\labelsep\usecounter{enumi}}}
 \relax
\def\reflist{\section*{References\markboth
        {REFLIST}{REFLIST}}\list
        {[\arabic{enumi}]\hfill}{\settowidth\labelwidth{[999]}
        \leftmargin\labelwidth
        \advance\leftmargin\labelsep\usecounter{enumi}}}
 \relax
\makeatletter
\newcounter{pubctr}
\def\publist{\@ifnextchar[{\@publist}{\@@publist}}
\def\@publist[#1]{\list
        {[\arabic{pubctr}]\hfill}{\settowidth\labelwidth{[999]}
        \leftmargin\labelwidth
        \advance\leftmargin\labelsep
        \@nmbrlisttrue\def\@listctr{pubctr}
        \setcounter{pubctr}{#1}\addtocounter{pubctr}{-1}}}
\def\@@publist{\list
        {[\arabic{pubctr}]\hfill}{\settowidth\labelwidth{[999]}
        \leftmargin\labelwidth
        \advance\leftmargin\labelsep
        \@nmbrlisttrue\def\@listctr{pubctr}}}
 \relax
\makeatother
\newskip\humongous \humongous=0pt plus 1000pt minus 1000pt

\newif\ifdtup

\relax

\def\be{\begin{equation}}
\def\ee{\end{equation}}
\def\ba{\begin{eqnarray}}
\def\ea{\end{eqnarray}}

\def\no{\noindent}

\def\bl{\bigl}
\def\br{\bigr}

\def\IR{\relax{\rm I\kern-.18em R}}

\def\0{{\sst{(0)}}}
\def\1{{\sst{(1)}}}
\def\2{{\sst{(2)}}}
\def\3{{\sst{(3)}}}
\def\4{{\sst{(4)}}}
\def\5{{\sst{(5)}}}
\def\6{{\sst{(6)}}}
\def\7{{\sst{(7)}}}
\def\8{{\sst{(8)}}}
\def\n{{\sst{(n)}}}

 \let\m=\mu \let\n=\nu  \let\r=\rho
 \let\t=\tau

 \def\bd{\begin{document}} \def\ed{\end{document}}
\def\ds{\documentstyle} \let\fr=\frac \let\bl=\bigl \let\br=\bigr
\let\Br=\Bigr \let\Bl=\Bigl
\let\bm=\bibitem
\let\na=\nabla
\let\pa=\partial \let\ov=\overline
\def\ft#1#2{{\textstyle{{\scriptstyle #1}\over {\scriptstyle #2}}}}
\def\fft#1#2{{#1 \over #2}}
\def\del{\partial}
\def\sst#1{{\scriptscriptstyle #1}}
 \def\oneone{\rlap 1\mkern4mu{\rm l}}
\def\ie{{\it i.e.\ }}
\def\via{{\it via}}
\def\semi{{\ltimes}}
\def\str{{\rm str}}
\def\Dm{{{D_{\sst{max}}}}}
\def\vac{ \left | 0 \right \rangle }
\def\kvac{ \left | k \right \rangle }

\def\sp{\; \; \;}

\def\bol{ \left | B (p^+) \right \rangle}
\def\bo1{ \left | B^0 (p^+) \right \rangle}

\def\bolt{ \left | B (p^+) \right \rangle_{\t}}

\def\boxl{ \left | B (x^-) \right \rangle}

\def\<{ \langle }
\def\>{ \rangle }

\def\vf{\varphi}

\def\ls{{(l,0)}}
\def\lv{{(l,\pm1)}}
\def\lt{{(l,\pm2)}}

\def\lse#1{{(l_{#1},0)}}
\def\lve#1{{(l_{#1},\pm1)}}
\def\lte#1{{(l_{#1},\pm2)}}

\def\lsg#1{{5(l_{#1},0)}}
\def\lvg#1{{5(l_{#1},\pm1)}}
\def\ltg#1{{5(l_{#1},\pm2)}}

\def\lsi#1{{5{(#1,0)}}}
\def\lvi#1{{5{(#1,\pm1)}}}
\def\lti#1{{5{(#1,\pm2)}}}

\def\lsr#1{{1{(#1,0)}}}
\def\lvr#1{{1{(#1,\pm1)}}}
\def\ltr#1{{1{(#1,\pm2)}}}

\def\cn{{\cal N}}
\def\cao{{\cal O}}
\def\cD{{\cal D}}
\def\cE{{\cal E}}
\def\cF{{\cal F}}
\def\cG{{\cal G}}
\def\cH{{\cal H}}
\def\cK{{\cal K}}
\def\cO{{\cal O}}
\def\cP{{\cal P}}
\def\cQ{{\cal Q}}
\def\cR{{\cal R}}
\def\cS{{\cal S}}
\def\cT{{\cal T}}
\def\cU{{\cal U}}
\def\cV{{\cal V}}
\def\cW{{\cal W}}

\newcommand{\nono}{\nonumber}
\newcommand{\dtilde}[1]{\tilde{\tilde{#1}}}
\newcommand{\hatb}[1]{\hat{\ov{#1}}}
\newcommand{\hatt}[1]{\hat{\tilde{#1}}}
\newcommand{\emnr}{{e_\m}^{\n\r}}

\begin{document}
\renewcommand{\theequation}{\arabic{equation}}

\newcommand{\beq}{\begin{equation}}
\newcommand{\eeq}[1]{\label{#1}\end{equation}}
\newcommand{\ber}{\begin{eqnarray}}
\newcommand{\eer}[1]{\label{#1}\end{eqnarray}}
\newcommand{\eqn}[1]{(\ref{#1})}
\begin{titlepage}
\begin{center}

\hfill May 2016\\

\vskip 1in

{\large \bf Aspects of non-associative structures in physics}\footnote{Contribution to the
{\em Workshop on Non-commutative Field Theory and Gravity}, 21--27 September 2015,
Corfu, Greece; to appear in the Proceedings of Science. Also, based on a lecture delivered at the
{\em Workshop on Quantized Geometry and Physics}, 23--26 May 2014, Bayrischzell, Germany and at the
{\em Joint ERC Workshop on MassTeV, Superfields and Strings \& Gravity}, 16--18 October 2013, Munich,
Germany.}

\vskip 0.6in

{\bf Ioannis Bakas}
\vskip 0.2in
{\em Department of Physics, School of Applied Mathematics and Physical Sciences \\
National Technical University, 15780 Athens, Greece\\
\vskip 0.2in
\footnotesize{\tt bakas@mail.ntua.gr}}\\

\end{center}

\vskip .8in

\centerline{\bf Abstract}

\no
We summarize the emergence of non-commutative/non-associative structures in Dirac's generalization of
Maxwell theory, focusing mostly on the magnetic field analogue of the non-geometric R-flux string model.
The cohomological interpretation of the obstructions to associativity in terms of 3-cocycles and the use
of the star product as alternative to ordinary quantization are also discussed in this context.

\vfill
\end{titlepage}
\eject

\def\baselinestretch{1.2}
\baselineskip 16 pt
\noindent



{\bf 1. Introduction:} It was established in recent years that non-geometric closed string backgrounds
exhibit non-commutativity/non-associativity among their coordinates. The prime example is provided by
the so called R-flux model obtained by a sequence of T-dualities along the coordinates of a torus
$T^3$ with constant 3-form flux. In that case, the commutation relations among the coordinates and
momenta assume the following form,
\be
[ x^i, ~ x^j ] = i  R \, \epsilon^{ijk} p_k , ~~~~
[x^i, ~ p^j] = i \, \delta^{ij}, ~~~~ [p^i, ~ p^j] = 0 \, ,
\ee
where $R$ is a constant provided by the 3-form flux in appropriate units. As a result, the Jacobiator
of the string coordinates does not vanish,
\be
[ x^1, ~ x^2, ~ x^3] = [[ x^1, ~ x^2], ~ x^3]+{\rm cyclic ~ permutations} = -3R \, ,
\ee
giving rise to non-associativity. There are more examples of non-geometric closed strings exhibiting
non-commutative/non-associative structures, but they are more complicated lying beyond the scope of the
present exposition.

There is a parallel story in Dirac's generalization of Maxwell theory in the presence of magnetic
sources. One often considers a point particle in the field of a single monopole but it is also legitimate
consider the motion of the particle in the field of a continuous distribution of magnetic charge. In
that case, non-commutativity/non-associativity arises in momentum space, hereby posing a problem in the
quantization of the system (under the assumption that magnetic monopoles are for real).

In the following, we focus on the magnetic field analogue of the R-flux string
model which serves as example to discuss the emergence of non-commutativity/non-associativity
together with its cohomological interpretation and the use of star product as alternative to quantization.
The presentation is based on the material contained in our earlier work on the subject:
I. Bakas and D. L\"ust, ``3-cocycles, non-associative star products and the magnetic paradigm
of R-flux string vacua", JHEP \underline{1401} (2014) 171, arXiv:1309.3172 [hep-th]
and references therein.

{\bf 2. Non-associativity in the presence of magnetic sources:} A spinless point particle with
electric charge $e$ and mass $m$ placed in the magnetic field background $\vec{B}(\vec{x})$
has the following commutation relations among its coordinates and momenta (in units $\hbar = 1$),
\be
[x^i , ~ p^j] = i \, \delta^{ij} , ~~~~~
[x^i , ~ x^j] = 0 , ~~~~~ [p^i , ~ p^j] = ie \, \epsilon^{ijk} B_k (\vec{x}) ,
\ee
leading to non-commutativity in momentum space in the context of Maxwell theory.
In Dirac's generalization of Maxwell theory, we have  $\vec{\nabla} \cdot \vec{B} \neq 0$ in the
presence of magnetic sources, and, thus, associativity is also lost in momentum space, since
\be
[p^i , ~ p^j , ~ p^k] = [[p^i , ~ p^j] , ~ p^k] + {\rm cyclic ~ permutations} =  - e ~\epsilon^{ijk}
\vec{\nabla} \cdot \vec{B}  \neq  0 .
\ee
This provides a simple model for non-commutativity/non-associativity though it arises in momentum rather than
configuration space.

We consider a continuous spherically symmetric distribution of magnetic
charge in space, $\rho (x)$, to study some of the implications of non-commutativity/non-associativity
in classical and quantum theory. Setting $x^2 = \vec{x} \cdot \vec{x}$, we have
\be
\vec{\nabla} \cdot \vec{B} = \rho (x) .
\ee
The particular solution of the inhomogeneous equation is expressed as
\be
\vec{B} (\vec{x}) = {\vec{x} \over f(x)}, ~~~~~~ \rho (x) = {3f(x) - x f^{\prime} (x) \over f^2 (x)} .
\ee
It is a consistent solution of Dirac's generalization of Maxwell theory in the limit of static sources,
since $\vec{\nabla} \times \vec{B} = 0$.

Using the Hamiltonian $H = \vec{p} \cdot \vec{p}/2m$, the Lorentz force acting on the
point particle in the magnetic field background is
\be
{d \vec{p} \over dt} = i \, [H , ~ \vec{p} ~] = {e \over 2m} \, (\vec{p} \times \vec{B} - \vec{B} \times \vec{p} ~) ,
\ee
which for $\vec{B}(\vec{x}) = \vec{x} / f(x)$ takes the special form
\be
m \, {d^2 \vec{x} \over dt^2} = - {e \over f(x)} \left(\vec{x} \times {d \vec{x} \over dt} \right) .
\ee
The Lorentz force is proportional to angular momentum and does no work. However, the equations are not integrable
because the angular momentum of the point particle is not conserved, in general,
\be
{d \over dt} \left(m ~ \vec{x} \times {d \vec{x} \over dt} \right) = {e ~ x^3 \over f(x)} ~ {d \hat{x} \over dt} .
\ee
We conclude that non-associativity accounts for the breakdown of angular symmetry.

The only exception to the general rule is the Dirac monopole with magnetic charge $g$, having $f(x) = x^3/g$,
so that $\rho (x) = 4 \pi g ~ \delta (x)$. In this case, the celebrated Poincar\'e vector
\be
\vec{J} = m \, \vec{x} \times {d \vec{x} \over dt} - e g ~ \hat{x}
\ee
provides the improved angular momentum of the particle that is conserved. Also,
the apparent violation of non-associativity in a Dirac monopole field is localized to a point and it can
be eliminated by imposing the boundary condition $\Psi(0) =0$ on the wave-functions of the system.
Finite translations in space also associate when Dirac's quantization condition
$eg = n \in \mathbb{Z}$  ($ \times \hbar /2$) is satisfied. In all other cases, non-associativity is for
real, obstructing canonical quantization.

Another notable example is provided by the choice  $f(x) = 3/\rho$ so that $\rho (x) = \rho$ is constant and
$\vec{B}(\vec{x}) = \rho \, \vec{x} /3$. As such, it provides the magnetic field analogue of the R-flux string model.
It is a genuinely non-commutative/non-associative model that will occupy the rest of this study.
It is the simplest magnetic background obtained by homogeneous distribution of magnetic charge all over space.

{\bf 3. Cohomological characterization of non-associativity:} Focusing to the case of constant
magnetic charge density, which is the magnetic field analogue of the R-flux string
model, we go on to characterize the emergence of non-associativity in terms of Lie algebra cohomology.
The basic commutation relations take the following form (in units $\hbar = 1$),
\be
[ p^i, ~ p^j ] = iR \, \epsilon^{ijk} x_k , ~~~~
[x^i, ~ p^j] = i \, \delta^{ij}, ~~~~ [x^i, ~ x^j] = 0
\ee
with parameter $R = e \rho/3$. In this case, the Jacobiator among the momenta does not vanish,
\be
[ p^1, ~ p^2, ~ p^3] ~ = ~ [[ p^1, ~ p^2], ~ p^3]+{\rm cyclic ~ permutations} = -3R ,
\ee
signaling the breakdown of associativity all over space and not just at a point.

The obstruction to non-associativity is a 3-cocycle in the cohomology theory of the Abelian
Lie algebra ${\bf t}_6$ associated to translations in phase space. Letting $T_I = (x^i,  \, p^i)$
be the generators of ${\bf t}_6$, we choose a 3-cochain $c_3(T_I, T_J, T_K) = 0$
with $c_3(p^1, p^2, p^3) = 1$, up to normalization, and $c_3(T_I, T_J, T_K) = 0$ for all
other choices of generators (i.e., when at least one $T$ is $x$). We have, in particular,
\be
[T_I, ~ T_J , ~ T_K] \sim c_3 (T_I, T_J, T_K)
\ee
and, thus, only the Jacobiator $[p^1, p^2, p^3]$ does not vanish.
The obstruction satisfies the 3-cocycle condition $dc_3 (T_I, T_J, T_K, T_L) = 0$,
since for any four elements of ${\bf t}_6$ we have
\ba
& & c_3([T_I , T_J], ~ T_K, ~ T_L) - c_3([T_I , T_K], ~ T_J, ~ T_L) + c_3([T_I , T_L], ~ T_J, ~ T_K)  +
\nonumber\\
& & c_3([T_J , T_K], ~ T_I, ~ T_L) - c_3([T_J , T_L], ~ T_I, ~ T_K) + c_3([T_K , T_L], ~ T_I, ~ T_J) = 0 .
\ea

Alternatively, we can describe the obstruction to associativity in terms of the Abelian group of
translations in phase space. For this, we exponentiate the action of the position and momentum generators.
The corresponding group elements are
\be
U(\vec{a}, ~ \vec{b}) ~ = ~ e^{i(\vec{a} \cdot \vec{x} + \vec{b} \cdot \vec{p})} ,
\ee
satisfying the composition law
\be
U(\vec{a}_1 , \vec{b}_1) U (\vec{a}_2 , \vec{b}_2) ~ = ~
e^{-{i \over 2} (\vec{a}_1 \cdot \vec{b}_2 - \vec{a}_2 \cdot \vec{b}_1)} ~
e^{-i {R \over 2} (\vec{b}_1 \times \vec{b}_2) \cdot \vec{x}} ~~
U (\vec{a}_1 + \vec{a}_2, ~ \vec{b}_1 + \vec{b}_2) .
\ee
Successive composition of any three group elements $U_i = U(\vec{a}_i, \vec{b}_i)$ yields
\be
(U_1 ~ U_2) ~
U_3 ~ = ~ e^{-i {R \over 2} (\vec{b}_1 \times \vec{b}_2) \cdot \vec{b}_3} ~
U_1 ~ (U_2 ~ U_3) .
\ee

If $R$ were zero, a projective representation of the Abelian group of translations would be in place. The phase factor
\be
\varphi_2 (\vec{a}_1 , \vec{b}_1 ; \vec{a}_2 , \vec{b}_2) = \vec{a}_1 \cdot \vec{b}_2 - \vec{a}_2 \cdot \vec{b}_1
\ee
is a real-valued 2-cocycle in group cohomology, satisfying
\be
d \varphi_2 (\vec{b}_1 , \vec{b}_2 , \vec{b}_3) \equiv
\varphi_2 (\vec{b}_2, \vec{b}_3) - \varphi_2 (\vec{b}_1 + \vec{b}_2 , \vec{b}_3) + \varphi_2 (\vec{b}_1 , \vec{b}_2 + \vec{b}_3) -
\varphi_2 (\vec{b}_1 , \vec{b}_2) = 0
\ee
and, thus, the associator is inert to it, as in ordinary quantum mechanics.

When $R \neq 0$, as in our case, there is an additional $x$-dependent factor in the composition
law of the group elements, which gives rise to a phase in the associator of any three group elements,
\be
\varphi_3 (\vec{b}_1 , \vec{b}_2 , \vec{b}_2)  = (\vec{b}_1 \times \vec{b}_2) \cdot \vec{b}_3 .
\ee
This phase is a real-valued 3-cocycle in the cohomology of the
Abelian group of translations in phase space, satisfying the condition
\ba
& & d \varphi_3 (\vec{b}_1 , \vec{b}_2 , \vec{b}_3 , \vec{b}_4) ~ \equiv ~
\varphi_3 (\vec{b}_2 , \vec{b}_3 , \vec{b}_4) - \varphi_3 (\vec{b}_1 + \vec{b}_2 , \vec{b}_3 , \vec{b}_4) +
\nonumber\\
& & \varphi_3 (\vec{b}_1 , \vec{b}_2 + \vec{b}_3 , \vec{b}_4) - \varphi_3 (\vec{b}_1 , \vec{b}_2 , \vec{b}_3 + \vec{b}_4) +
\varphi_3 (\vec{b}_1 , \vec{b}_2 , \vec{b}_3) ~ = ~ 0 ~.
\ea
A schematic representation is provided by Mac Lane's pentagon relating the composition of four group elements
$(U_1 U_2)(U_3 U_4)$ to $U_1 (U_2 (U_3 U_4))$ to $((U_1 U_2) U_3)U_4$ to
$U_1 ((U_2 U_3)U_4)$ to $(U_1 (U_2 U_3)) U_4$.

{\bf 4. Star product as alternative to quantization:} When $R=0$, all classical observables $f(x, p)$ on
phase space are assigned to operators $\hat{F}(\hat{x}, \hat{p})$ acting on Hilbert space ${\cal H}$.
Their product is non-commutative but associative.
An equivalent description is provided by Moyal star-product in phase space. We Fourier analyze
\be
f(\vec{x}, \vec{p}) = { 1 \over (2 \pi)^3} \int d^3 a d^3 b ~ \tilde{f} (\vec{a}, \vec{b})
e^{i(\vec{a} \cdot \vec{x} + \vec{b} \cdot \vec{p})}
\ee
and apply Weyl's correspondence rule to assign self-adjoint operators
\be
\hat{F}(\hat{\vec{x}}, \hat{\vec{p}}) = { 1 \over (2 \pi)^3} \int d^3 a d^3 b ~ \tilde{f} (\vec{a}, \vec{b})
\hat{U} (\vec{a} , \vec{b}) ,
\ee
where
\be
\hat{U}(\vec{a}, ~ \vec{b}) = e^{i(\vec{a} \cdot \hat{\vec{x}} + \vec{b} \cdot \hat{\vec{p}})} .
\ee

The product of any two operators takes the following form,
\be
\hat{F}_1 \cdot \hat{F}_2 = { 1 \over (2 \pi)^{6}} \int d^3 a_1 d^3 b_1 d^3 a_2 d^3 b_2 ~ \tilde{f}_1 (\vec{a}_1, \vec{b}_1)
\tilde{f}_2 (\vec{a}_2, \vec{b}_2) \hat{U} (\vec{a}_1, ~ \vec{b}_1) \hat{U} (\vec{a}_2, ~ \vec{b}_2) ,
\ee
which can be subsequently worked out using the group product composition law
\be
\hat{U}(\vec{a}_1, ~ \vec{b}_1) \hat{U}(\vec{a}_2, ~ \vec{b}_2) =
e^{-{i \over 2} (\vec{a}_1 \cdot \vec{b}_2 - \vec{a}_2 \cdot \vec{b}_1)}
\hat{U} (\vec{a}_1 + \vec{a}_2, ~ \vec{b}_1 + \vec{b}_2 ) .
\ee
The 2-cocycle of the translation group $\varphi_2 (\vec{a}_1, \vec{b}_1; \vec{a}_1, \vec{b}_1) =
\vec{a}_1 \cdot \vec{b}_2 - \vec{a}_2 \cdot \vec{b}_1$ makes the product of the
corresponding phase space functions non-commutative but associative.
The result turns out to be
\be
(f_1 \star f_2) (\vec{x}, \vec{p}) = e^{{i \over 2} \left(\vec{\nabla}_{x_1} \cdot \vec{\nabla}_{p_2} -
\vec{\nabla}_{x_2} \cdot \vec{\nabla}_{p_1} \right)} f_1 (\vec{x}_1, \vec{p}_1)
f_2 (\vec{x}_2, \vec{p}_2) |_{\vec{x}_1 = \vec{x}_2 = \vec{x}; ~ \vec{p}_1 = \vec{p}_2 = \vec{p}} \, ,
\ee
giving rise to the series expansion
\be
(f_1 \star f_2) (\vec{x}, \vec{p}) = (f_1 \cdot f_2)(\vec{x}, \vec{p}) + {i \over 2} \{f_1 , ~ f_2 \} + \cdots ~.
\ee

The usual product of functions is deformed by derivative terms, already seen in the first correction
provided by the Poisson bracket, leading to non-commutative geometry in phase space as the notion of the point
becomes fuzzy. The deformation parameter is Planck's constant $\hbar$ which is not seen here as it is normalized
to 1. Then, in this context, quantum dynamics is equivalently described by the Moyal bracket
\be
\{\{f_1 , ~ f_2 \}\} \equiv -i (f_1 \star f_2 - f_2 \star f_1) = \{f_1 , ~ f_2 \} + {\rm higher ~ derivatives}
\ee
that deforms the Poisson bracket by higher derivative terms and it acts as derivation
\be
\{\{f_1 , ~ f_2 \star f_3 \}\} = f_2 \star \{\{f_1 , ~ f_3 \}\} + \{\{f_1 , ~ f_2 \}\} \star f_3 .
\ee

When $R \neq 0$, the rules of canonical quantization do not apply, but it is
still possible to define a non-commutative/non-associative star-product.
We follow the same line of thought as before, assigning to $f(\vec{x}, \vec{p})$
\be
F(\vec{x}, \vec{p}) = { 1 \over (2 \pi)^3} \int d^3 a d^3 b ~ \tilde{f} (\vec{a}, \vec{b})
U (\vec{a} , \vec{b})
\ee
and using the generalized composition law
\be
U(\vec{a}_1 , \vec{b}_1) U (\vec{a}_2 , \vec{b}_2) =
e^{-{i \over 2} (\vec{a}_1 \cdot \vec{b}_2 - \vec{a}_2 \cdot \vec{b}_1)} ~
e^{-i {R \over 2} (\vec{b}_1 \times \vec{b}_2) \cdot \vec{x}} ~~
U (\vec{a}_1 + \vec{a}_2, ~ \vec{b}_1 + \vec{b}_2) .
\ee
The result is the non-commutative/non-associative $x$-dependent star-product
\ba
(f_1 \star_x f_2) (\vec{x}, \vec{p}) & = & e^{i {R \over 2} ~ \vec{x} \cdot (\vec{\nabla}_{p_1} \times \vec{\nabla}_{p_2})}
e^{{i \over 2} \left(\vec{\nabla}_{x_1} \cdot \vec{\nabla}_{p_2} -
\vec{\nabla}_{x_2} \cdot \vec{\nabla}_{p_1} \right)}
\nonumber\\
& & f_1 (\vec{x}_1, \vec{p}_1)
f_2 (\vec{x}_2, \vec{p}_2) |_{\vec{x}_1 = \vec{x}_2 = \vec{x}; ~ \vec{p}_1 = \vec{p}_2 = \vec{p}} ~.
\ea

In this case, there are no operators assigned to the classical observables $f(\vec{x}, \vec{p})$, since the
association to $F(\vec{x}, \vec{p})$ is only formal. The point of view we adopt here is that the star product
is still a viable operation that substitutes the notion of quantization. Then, in this context, quantum
dynamics is formulated solely in terms of the bracket
\be
\{\{f_1 , ~ f_2 \}\}_x \equiv -i (f_1 \star_x f_2 - f_2 \star_x f_1) ,
\ee
providing a non-associative generalization of the Moyal bracket. It does not act as derivation, since
\be
\{\{f_1 , ~ f_2 \star_x f_3 \}\}_x \neq f_2 \star_x \{\{f_1 , ~ f_3 \}\}_x + \{\{f_1 , ~ f_2 \}\}_x \star_x f_3
\ee
and the Jacobiator does not vanish. We have, in particular,
\be
\{\{f_1 (p), ~ f_2 (p), ~  f_3 (p) \}\}_x \neq 0 .
\ee

When $R \sim e \rho$ is not constant, the construction of the star product is technically more involved
and it will not be discussed. We only note here that the group product law $U(\vec{a}_1 , \vec{b}_1) U (\vec{a}_2 , \vec{b}_2)$
can not be found in closed form for general $\rho (x)$.

{\bf 5. Conclusions and discussion:} Motivated by the emergence of non-commutative and non-associative structures
in non-geometric closed sting models, we presented the paradigm of magnetic field backgrounds in Maxwell-Dirac
theory that exhibit similar structures. We focused mostly to the case of constant magnetic charge density,
which is the analogue of the R-flux string model, and discussed the cohomological interpretation of the obstructions
to associativity in terms of 3-cocycles and the use of the star product as a viable alternative to ordinary quantization of a
spinless electrically charged point particle.

In view of possible generalizations to other backgrounds, it is interesting to establish
a dictionary between non-geometric string vacua and distributions of magnetic charge in Maxwell-Dirac theory.
The simplest problem is to find the string analogue of the Dirac monopole, and, then, develop the
dictionary to find the string analogue of distributing magnetic charge all over space.
According to our discussion, non-associativity is distributed and not localized in either case.

\vspace{0.5cm}

{\bf Acknowledgments:} I thank the organizers for their kind invitation to present
this work and for the stimulating atmosphere they have created during the conference.

\end{document}